\documentstyle[12pt,psfig]{article}
\begin{document}
\thispagestyle{empty}
\noindent
\begin{flushright}
        OHSTPY-HEP-T-98-009\\
        April 1998
\end{flushright}

\vspace{1cm}
\begin{center}
  \begin{Large}
  \begin{bf}
   Dynamical SUSY Breaking with a Hybrid Messenger Sector
   \\
  \end{bf}
 \end{Large}
\end{center}
  \vspace{1cm}
 
    \begin{center}
    Stuart Raby$^\dagger$ and Kazuhiro Tobe$^*$\\
      \vspace{0.3cm}
\begin{it}
Department of Physics, \\
The Ohio State University, \\
174 W. 18th Ave., \\
Columbus, Ohio  43210\\
$^\dagger$raby@mps.ohio-state.edu \\ $^*$tobe@pacific.mps.ohio-state.edu\\
\end{it}

  \end{center}
  \vspace{1cm}
\centerline{\bf Abstract}
\begin{quotation}
\noindent
In this paper we present a dynamical model of SUSY breaking with a hybrid
messenger sector.  SUSY is broken dynamically at a scale of order $10^9$ GeV 
via strong SU(2) gauge interactions.  SUSY breaking is then transmitted to 
the observable sector via two distinct sources: (1)  messengers, carrying
Standard Model gauge quantum numbers, with the messenger mass of order
 $10^{15}$ GeV, and (2)
 the D term of an anomalous U(1)$_X$.  The model is quite constrained.
The messenger scale is fixed by the Fayet-Iliopoulos term for the anomalous 
U(1)$_X$ interaction.   In addition, we show that the D term SUSY breaking 
contributions to squark and slepton masses are  "naturally" the same order 
as those coming from the messengers.

\end{quotation}
\vfill\eject 

\section{Introduction}
 
Supersymmetry [SUSY] is a strongly motivated candidate for new physics
beyond the Standard Model [SM].  The minimal supersymmetric particle content is 
well defined and the interactions of all the new superparticles [sparticles]
are constrained
by the observed SM interactions as long as the theory has a conserved R parity.
One might naively expect that it would be a simple task to search for the new
sparticles.  Of course, it is not easy and the reason is two-fold.  First, 
the masses of the new superparticles depend on how supersymmetry  is 
broken and secondly, since these particles must be produced in high
energy/luminosity collisions, the signal to background ratio is generically quite
small.  Thus in order to find SUSY one must look for particular signatures which
may be pulled out of the SM background by skilled experimental hands.  In this
paper we present another scenario for obtaining squark, slepton and gaugino
masses.   The mechanism is a hybrid of gauge-mediated SUSY breaking via
messengers carrying SM gauge interactions and D-term SUSY breaking associated 
with an anomalous U(1)$_X$ gauge symmetry.  Similar ideas have recently been 
discussed in the literature\cite{dvali,binetruy,dilaton,farragi}.  The new feature 
in this paper is that the SUSY breaking contributions of the messengers and
that of the D term are ``naturally" expected to be of the same order of 
magnitude.  In addition the messenger scale is determined by the
Fayet-Illiopoulos D term contribution to U(1)$_X$.  We discuss the spectrum 
of sparticle masses for this theory.    Note, in recent years, anomalous U(1)$_X$ 
symmetries have also been used to construct models of fermion and sfermion 
masses\cite{ibanez} - \cite{nelson}.

\section{Dynamical SUSY Breaking Sector}

Consider an SU(N) gauge group with N$_F$ = N flavors.   Such a theory
has a quantum moduli space first considered by Seiberg\cite{seiberg}.
Models of dynamical SUSY breaking using these theories as a starting point
have also been discussed in the literature\cite{intriligator,murayama}.  We
shall focus on the simplest model of this type with N = N$_F$ = 2.

The model includes the chiral superfields --- $Q_{i \alpha}, \bar Q^{i \alpha}$
--- transforming as a $2 + \bar 2 $ of the strong SU(2) gauge symmetry where i = 1,2
is a gauge index and $\alpha = 1,2$ is a flavor index.   The theory has an
SU(4) flavor symmetry, but we shall break this full flavor symmetry by weakly
gauging an SU(2)$_F$ subgroup such that $Q, \bar Q$ transform as a $2, \bar 2 $
of SU(2)$_F$, respectively.  In addition, the theory has a global baryon number
symmetry U(1)$_B$ where  ($Q, \bar Q$) have charge   (1, -1). It
will be convenient to work in terms of SU(2) strong gauge singlet superfields defined
by
\begin{eqnarray}
{M^\beta}_\alpha  & =  \bar Q^{i \beta} \; Q_{i \alpha} &  \nonumber \\
B_{\alpha \beta} & = {1 \over \sqrt{2}} \; Q_{i \alpha} \; Q_{j \beta} \;
\epsilon^{ij} & = - B_{\beta \alpha} \nonumber \\
\bar B^{\alpha \beta} & = {1 \over \sqrt{2}} \; \bar Q^{i \alpha} \; 
\bar Q^{j \beta} \; \epsilon_{ij} & = - \bar B^{\beta \alpha}
\end{eqnarray}
The quantum moduli space is given by the equation
\begin{eqnarray}  
det\, M - \bar B^{\alpha \beta} \; B_{\alpha \beta} & = \Lambda_s^4 & 
\end{eqnarray}
where $\Lambda_s$ is the dynamically determined scale of the strong
SU(2) gauge symmetry.

The dynamical SUSY breaking sector of the theory includes, in addition to
the SU(2)$\times$SU(2)$_F$ gauge theory with chiral states described above, 
the states $X,\;\; A^{\alpha \beta} = - A^{\beta \alpha},\;\; \bar A_{\alpha \beta}
= - \bar A_{\beta \alpha},$ and $ {S^\alpha}_\beta$ where $X, A, \bar A$ are 
SU(2)$\times$SU(2)$_F$ singlets and $ {S^\alpha}_\beta $ is an SU(2) singlet and
SU(2)$_F$ triplet.  They have charge (0, -2, 2, 0) under U(1)$_B$, respectively. 
Note,  we also can define the fields
$S_a,
\; a = 1,2,3$ by $   {S^\alpha}_\beta \equiv  \sqrt{2} \; S_a \; {(T_a)^\alpha}_\beta
$.

The most general superspace potential, invariant under
SU(2)$\times$SU(2)$_F$$\times$U(1)$_B$, for the dynamical SUSY breaking sector is
given by
\begin{eqnarray}  
W = & {\lambda \over \sqrt{2}} \; X \; Tr M + \lambda' \; {S^\alpha}_\beta
 \; {M^\beta}_\alpha &  \nonumber \\
  & + \lambda'' \; A^{\alpha \beta} \; B_{\alpha \beta} + \bar \lambda'' \;
  \bar A_{\alpha \beta} \; \bar B^{\alpha \beta} & \nonumber \\
  & + U [ det\, M - \bar B^{\alpha \beta} \; B_{\alpha \beta} - \Lambda_s^4]
  & \label{eq:suppot}
  \end{eqnarray}
  where  $U$ is a Lagrange multiplier enforcing quantum moduli space.
  
  This theory breaks supersymmetry dynamically.  We can study the theory
  for large values of the fields $X,\;\; A^{\alpha \beta},\;\; 
  \bar A_{\alpha \beta}, \;\; {S^\alpha}_\beta$ by integrating out 
  the states ${M^\beta}_\alpha,\;\; B_{\alpha \beta}, \;\; \bar B^{\alpha \beta}$
  and the auxiliary field $U$.\footnote{Actually, since SUSY is broken the
  auxiliary field for $U$ gets a nonvanishing vev.  If one minimizes the
  full potential we obtain higher order corrections in $\lambda^2, \; 
  \lambda'^2$ etc. to the leading order expression for $W$ below.}  
  We find the effective superspace potential
  
 \begin{eqnarray}
 W & = -2 \; \Lambda_s^2 \; \sqrt{\lambda^2 \; X^2/2 \; - \; \lambda'^2 \; Tr S^2/2\;
- \; \lambda''
 \; \bar \lambda'' \; A^{\alpha \beta} \bar A_{\alpha \beta}} &
 \end{eqnarray}
 and the resulting scalar potential $ V = |{\partial W \over \partial X}|^2 +
  Tr |{\partial W \over \partial {S^\alpha}_\beta }|^2 + |{\partial W \over \partial  A^{\alpha
  \beta}}|^2 + |{\partial W \over \partial \bar A_{\alpha \beta}}|^2 $  is
	  given by (see fig. 1)
 
  \begin{eqnarray}
  V & = \Lambda_s^4 \;{(\lambda^4 \; |X|^2 \; + \; \lambda'^4 \; |{S^\alpha}_\beta|^2
   + \; \lambda''^2
 \; \bar \lambda''^2 \; (|A^{\alpha \beta}|^2 + |\bar A_{\alpha \beta}|^2)
 \over |\lambda^2 \; X^2/2 \; - \; \lambda'^2 \; Tr S^2/2\; - \; \lambda''
 \; \bar \lambda'' \; A^{\alpha \beta} \bar A_{\alpha \beta}|} &
\label{scalar_pot}
 \end{eqnarray}

\begin{figure}
	\centerline{ \psfig{file=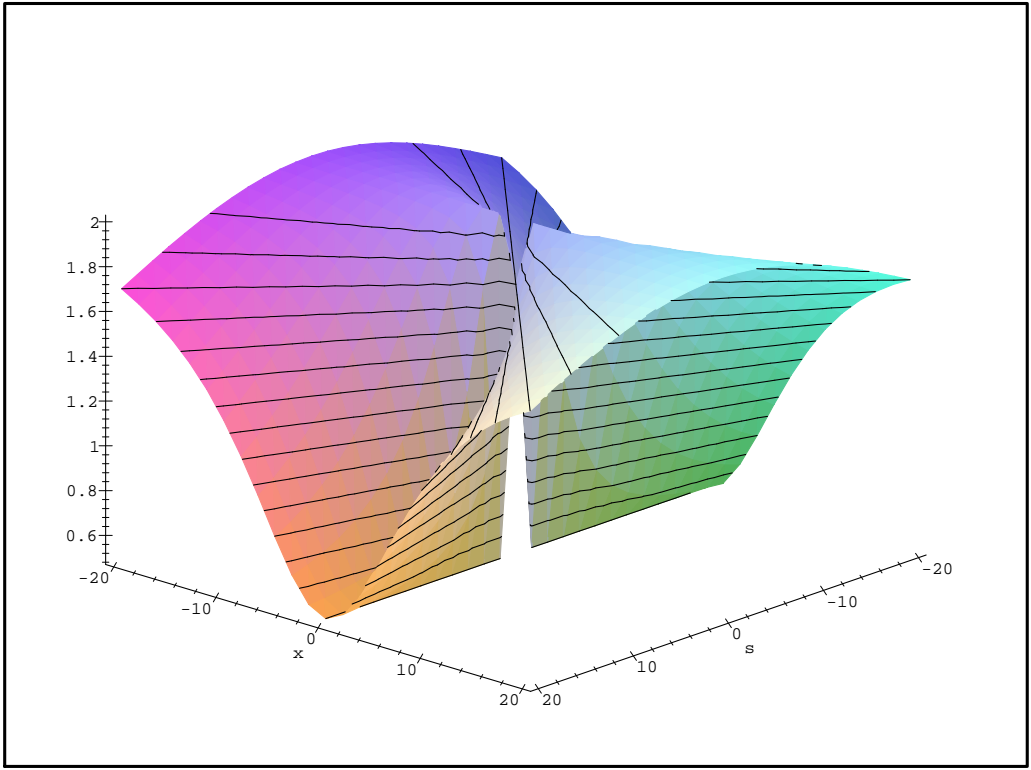,width=15cm,angle=-90}}
\caption{The scalar potential (eqn.~{\ref{scalar_pot}}).
Here we assume that $A={\bar A}=0$, $\lambda=1$, and $\lambda'=i/2$.}
\end{figure}

The potential has no supersymmetric minima.  There are several flat directions.
For example, if we take $\{X,\;\; A^{\alpha \beta},\;\; 
  \bar A_{\alpha \beta}, \;\; {S^\alpha}_\beta \} = \{ X, 0,0,0\}$ we have
  \begin{eqnarray}
   V(X,0,0,0) & = 2 \lambda^2 \; \Lambda_s^4 &
   \end{eqnarray}
  or  for  $\{ 0,0,0, S_3 \} $, we have
  \begin{eqnarray}
   V(0,0,0,S_3) & = 2 \lambda'^2 \; \Lambda_s^4 &
   \end{eqnarray}
   
For the range of parameters satisfying 
$\lambda' < \lambda \sim \lambda'' \sim \bar \lambda''$ the flat direction
with  $\{X,\;\; A^{\alpha \beta},\;\; 
  \bar A_{\alpha \beta}, \;\; {S^\alpha}_\beta \} = \{ 0,0,0, S_3 \}$
  is the lowest energy state.  Along this flat direction the
  superspace potential is given by 
  \begin{eqnarray} W & = - i \sqrt{2}\;  \lambda' \; S_3 \; \Lambda_s^2 &
  \end{eqnarray}
  
Radiative corrections lift this tree level degeneracy.   
As discussed in the literature\cite{murayama}, the Lagrangian receives
corrections when integrating out states between $S_3$ and $M$ ($M$ is either
the Planck scale or the string scale) a la Wilson.
We find
\begin{eqnarray}
{\cal L} & = \int d^4 \theta  \tilde Z_S |S_3|^2 
 + \int d^2 \theta W(S_3)  + h.c. &
 \end{eqnarray}
 where at one loop  
\begin{eqnarray} 
\tilde Z_S & = 1 + \gamma_S \ln(M^2/|S_3|^2) & \rm and  \\
 \gamma_S = & (2 \lambda'^2 - 4 g_F^2)/16 \pi^2 & \label{eq:gammaS1}
\end{eqnarray} is the anomalous
 dimension of $S_3$ and $g_F$ is the SU(2)$_F$ gauge coupling constant.  
 More generally we have
 \begin{eqnarray}
 \gamma_S &= - {1 \over 2} {\partial \ln \tilde Z_S \over \partial ln \mu} &
\label{eq:gamma}
 \end{eqnarray} where $\mu = |S_3|$ and both $\lambda'$ and $g_F$ are
 $\mu$ dependent.   Thus the scalar potential is given by
 \begin{eqnarray} V_0 & = K^{-1}_{S S^{\dagger}}\; |{\partial W \over \partial
 S_3}|^2 &
 \end{eqnarray} with $K(S, S^{\dagger}) =  \tilde Z_S |S_3|^2 $ and $K_{S
S^{\dagger}} \equiv  \partial^2 K(S, S^\dagger)/\partial S \partial S^\dagger$.  
Hence
 
 \begin{eqnarray} V_0 & \approx 2 \, \lambda'^2 \; \Lambda_s^4 / \tilde Z_S &
  \end{eqnarray}
 The extremum of the potential $\bar S_0$ is given by
  \begin{eqnarray}
  S_3 {\partial V_0 \over \partial S_3} & = {- 2 \, \lambda'^2 \; \Lambda_s^4 \over 
  \tilde Z_S} \; {\partial \ln \tilde Z_S \over \partial \ln |S_3|} & \nonumber \\
   &  =  2 V_0 \; \gamma_S(\bar S_0) & = 0  \label{eq:dVdS}
   \end{eqnarray}
This is satisfied for  $\gamma_S({\bar S_0}) \equiv 0$ or $\lambda'^2 = 2 g_F^2$.  

We can check that this is in fact a minimum. Consider the second derivative
\begin{eqnarray}
|S_3|^2 {\partial^2 V_0 \over \partial |S_3|^2} & = 2 V_0 \;
{\partial
\gamma_S(S_3) \over \partial \ln |S_3|} & \rm at \;\;S_3 = \bar S_0
\end{eqnarray}
 We may obtain $\gamma_S(S_3)$ in an expansion in
$\ln(|S_3|/|\bar S_0|)$.   Using the definitions of the beta functions
\begin{eqnarray} 
\beta_{\lambda'}  \equiv {\partial \lambda' \over \partial \ln \mu} & = \lambda'
(\gamma_S + \gamma_Q + \gamma_{\bar Q}) & \nonumber \\
\beta_{g_F}  \equiv {\partial g_F \over \partial \ln \mu} & = - b \, g_F^3/16 \pi^2
& \nonumber \\
\rm with  &  b = 3 C_2(SU(2)) - T(S) - 2 T(Q) - 2 T(\bar Q) & = 2 ,  \label{eq:b}
\end{eqnarray}
and eqn. \ref{eq:gammaS1} for $\gamma_S$, we find to first order in $\ln(|S_3|/|\bar
S_0|)$ 
\begin{eqnarray}
\gamma_S(S_3) & \approx \gamma_S(\bar S_0) + 2 \left({ \lambda'
\beta_{\lambda'}(\bar S_0)\over 8 \pi^2}
\; \ln(|S_3|/|\bar S_0|)  \; - \; {2 g_F \beta_{g_F}(\bar S_0) \over 8 \pi^2}\;
\ln(|S_3|/|\bar S_0|) \right)  & \rm or  \nonumber\\
\gamma_S(S_3) & \approx  2 \left({\lambda'^2(\bar S_0) \over 8 \pi^2} \; 
(\gamma_Q(\bar S_0) + \gamma_{\bar Q}(\bar S_0)) \; + \; 4 \ b \; ({ \alpha_F \over
4 \pi})^2 \right)
\; \ln({|S_3| \over |\bar S_0|}) &  \label{eq:gammaS}
 \end{eqnarray}
Thus
\begin{eqnarray}
{\partial
\gamma_S(S_3) \over \partial \ln |S_3|} & =  2 \left({\lambda'^2(\bar S_0) \over 8
\pi^2} \;  (\gamma_Q(\bar S_0) + \gamma_{\bar Q}(\bar S_0)) \; + \; 4 \ b \; ({
\alpha_F \over 4 \pi})^2 \right) & \nonumber \\
 &  \rm at \;\; S_3 = \bar S_0  &
 \end{eqnarray}
Now 
\begin{eqnarray}
\gamma_Q = \gamma_{\bar Q} & = {3 \over 2} ( \lambda'^2 - g_F^2 - g^2 )/16 \pi^2
\end{eqnarray}
and at  $\bar S_0$ we have $\lambda'^2 = 2 g_F^2$, hence
\begin{eqnarray}
\gamma_Q(\bar S_0) = \gamma_{\bar Q}(\bar S_0) & = {3 \over 2} (  g_F(\bar S_0)^2 -
g(\bar S_0)^2 )/16 \pi^2 .
\end{eqnarray}
Thus it is sufficient for  $g < g_F$ at $\bar S_0$ and $b  \ge 0$ for the second
derivative of $V_0$ to be positive.  
Although the SU(2) coupling
$g$ is necessarily large near $\Lambda_s$, and larger than the flavor coupling $g_F$,
it varies more quickly than $g_F$ and  can easily be less than $g_F$ at $\bar S_0$.
We have $\alpha(\bar S_0) = g(\bar S_0)^2/4 \pi = \pi/ (2 \ln(\bar
S_0/\Lambda_s))$.
Thus for $\Lambda_s \sim 10^9$ GeV, $\bar S_0 \sim 10^{14}$ GeV we find
$\alpha(\bar S_0) \sim .14$.   If $g < g_F$ at $\bar S_0$,  $g_F$ can still be
perturbative and since it varies more slowly than $g$ it only gets strong
at a scale $\Lambda_F << \Lambda_s$.
With these caveats, the scalar potential $V_0$ has a minimum at $S_3 = \bar S_0$
defined by $\gamma_S(\bar S_0) = 0$.

We assume that $\bar S_0$ is at an intermediate scale of order $10^{14 - 15}$
GeV.  The precise value is not important.    We also assume that
${S^\alpha}_\beta$ couples to messengers which carry SM gauge quantum
numbers.   In the minimal messenger model we would have a coupling of the form

\begin{eqnarray}
W & \supset  {1 \over M}\; Tr ({S^\alpha}_\beta)^2 \; \bar 5 \; 5 &
\end{eqnarray}
where  $ 5, \bar 5$ are messengers transforming as indicated in $SU(5)$
representations\footnote{Note, the messenger mass $M_{mess} = <S_3>^2/M$ and we
may define an effective SUSY breaking scale  $F = 2 F_{S_3} <S_3> / M$.  
The scale $M$ in the higher dimension operator drops out in the ratio 
$F/M_{mess} = 2 F_{S_3}/<S_3>  \equiv \Lambda \sim 10^5$ GeV, where
$\Lambda$ is the effective SUSY breaking scale in the observable sector.}.   Note, however, this is not necessarily a 
grand unified model,
since we only need to gauge the $SU(3) \times SU(2) \times U(1)$ subgroup of 
$SU(5)$.  

Messengers contribute masses to squarks and sleptons proportional to
their SM quantum numbers.   At two loops we have
\begin{eqnarray} \tilde m^2 = & 8 {|F_{S_3}|^2 \over <S_3>^2} \; \left[
\sum_{i=1}^3 C_i\left({\alpha_i(M_{mess}) \over 4 \pi}\right)^2\right] & 
  \label{eq:mtilde} \end{eqnarray}
where $<S_3> = \bar S_0$, 
$C_3 = {4 \over 3}$ for color triplets and zero for singlets, 
$C_2= {3 \over 4}$ for weak doublets and zero for singlets, and $C_1 = 
{3 \over 5}{\left(Y\over2\right)^2}$, with the ordinary hypercharge $Y$
normalized as $Q = T_3 + {1 \over 2} Y$  and $\alpha_1$,  GUT normalized.

  Gauginos obtain mass at one loop given by
\begin{eqnarray}
    M_i = &  {\alpha_i(M_{mess})\over4\pi} \Lambda & ( {\rm for}\;\; i =
1,2,3). 
\end{eqnarray}

\section{Adding an Anomalous U(1)$_X$ to the theory}

Now consider how the theory changes if we include an anomalous U(1)$_X$ gauge
interaction.  Such interactions are natural in strings.   The scalar potential
given by
\begin{eqnarray}
 V  & = V_0 \; + \; {1 \over 2} \; D_X^2  &
 \end{eqnarray} 
now includes the contribution of the auxiliary field $D_X$  where
\begin{eqnarray}
D_X & = g_X ( \sum_a \,  Q^X_a \; |\phi_a|^2 \; + \; \xi ) &  
\end{eqnarray}  and
\begin{eqnarray}
 \xi  & = \epsilon \; M_{Pl}^2  &
 \end{eqnarray}  where 
\begin{eqnarray}
 \epsilon  & =  {g_X^2 Tr {\bf Q^X} \over 192 \; \pi^2} ({\sqrt{2} M_{st} \over
 g_X \; M_{Pl}})^2 .  \label{eq:epsilon}
 \end{eqnarray} 
$\xi$ is the Fayet-Iliopoulos D term which is generated at one loop\cite{fidterm,fidterm2}.
 $M_{st}$ is the string scale associated with the scale of compactification
 and $M_{Pl} = 2.4 \times 10^{18}$ GeV is the reduced Planck scale. Note
 in weak coupling string theory $M_{st} = g_X M_{Pl}/\sqrt{2}$.
 
 We assume that $Tr {\bf Q^X} > 0$ and that  $Q^X_S < 0$ so that the D term
 minimization condition $D_X(\phi_a) = 0$ has the solution  $\phi_a = 0$ for all
 $a$ EXCEPT for $S_3 \equiv S_0$ with $S_0 = \sqrt{\xi/|Q^X_S|}$.  Note, in
 general, $S_0 \neq \bar S_0$ discussed earlier.  Thus the true minimum of
 $V$ is neither at $S_0$ nor at $\bar S_0$ and at the minimum there will be
 a non-zero vacuum value for $D_X$.    We now show that the D term contribution to 
 sparticle masses is comparable to the SUSY breaking contribution from
 the messenger sector as discussed above.   Moreover the D term also sets
 the messenger scale.  This is the main result of the
 paper.\footnote{Note, once the anomalous U(1)$_X$ is included 
the fields \{$X,\;\; A^{\alpha \beta},\;\; 
  \bar A_{\alpha \beta}$\} are no longer needed to obtain SUSY breaking,
as discussed previously by Binetruy and Dudas\cite{binetruy}.  In addition,
 the presence of $\Lambda_s$ in the superspace potential (eqn.
\ref{eq:suppot}) explicitly breaks U(1)$_X$.  In string theories this
problem is resolved by the ``dilaton."  $\Lambda_s$ transforms as $det
M$ under U(1)$_X$ due to its implicit dependence on the ``dilaton" field $S = 1/g^2 + i
\, a$ where $a \rightarrow a + {1 \over 2} \delta _{GS}\theta$ (with $\delta_{GS} =
{Tr {\bf Q^X} \over 192 \pi^2}$) under a  U(1)$_X$ phase rotation by an angle
$\theta$\cite{binetruy,dilaton}.  Once the ``dilaton" is included the
anomally for U(1)$_X$ is cancelled a la Green-Schwartz.}
 
 The D term introduces the largest amount of curvature in the potential for
 $S_3$.   Hence  let $<S_3> =  S_0 +  \delta S$ where $\delta S$ is a small
 correction.   By minimizing $V$ and treating $\delta S$ perturbatively, we find
 
 \begin{eqnarray}
\delta S & =  - {1 \over 4 (Q^X_S)^2 \; g_X^2 \; S_0^3} \; (S_0 \; {\partial V_0 \over \partial
S_3}|_{S_3 = S_0})
 \end{eqnarray}
 and a shift in $D_X$ given by
 \begin{eqnarray}
g_X <D_X> & =  - 2 g_X^2 \; |Q^X_S| \; S_0 \; \delta S
 \end{eqnarray} 
 
 The SUSY breaking correction to scalar masses from the D term is given
 by
 \begin{eqnarray}
\delta \tilde m_a^2 & = g_X \; <D_X> \; Q^X_a &  
 \end{eqnarray}
 Using  the result for $(S_0 \; {\partial V_0 \over \partial
S_3}|_{S_3 = S_0})$ (eqn. \ref{eq:dVdS}) and  $V_0(S_0) = |F_{S_3}|^2$, 
we have

 \begin{eqnarray}
\delta \tilde m_a^2 & =  Q^X_a \; {1 \over |Q^X_S| \; S_0^2} \; V_0(S_0) \; \gamma_S(S_0) & \nonumber
\\   & = (Q^X_a / |Q^X_S|) \ {|F_{S_3}|^2 \over  S_0^2} \; \gamma_S(S_0) &
\label{eq:dtermmass}
 \end{eqnarray}

 Note, as discussed previously, $\gamma_S({\bar S_0}) \equiv 0$ and in eqn.
\ref{eq:gammaS} we have obtained $\gamma_S(S_3)$ in an expansion in
$\ln(|S_3|/|\bar S_0|)$.  
Hence $\gamma_S(S_0)$ is fourth order in gauge and Yukawa couplings.
Thus the D term contribution to squark and slepton masses (eqn.
 \ref{eq:dtermmass}) is naturally the same
 order as the contribution from messengers given in (eqn. \ref{eq:mtilde}),
 with $<S_3> = S_0$.   The sign of the mass correction is determined by the sign of
($Q^X_a \ \gamma_S(S_0)$).  Recall (eqn. \ref{eq:gamma}),  $\gamma_S =  - {1 \over 2}
{\partial \ln \tilde Z_S \over \partial ln \mu} $, hence 
\begin{eqnarray}
\gamma_S(S_0) & =  - {1 \over 2}\,|\bar S_0|^2{\partial^2 \ln \tilde Z_S \over 
\partial |S_0|^2} \ln ({|S_0| \over |\bar S_0|}) 
\end{eqnarray}  and the coefficient of the logarithm is simply related to the
second derivative of the scalar potential $V_0$ evaluated at the extremum $\bar
S_0$.  Thus at the minimum of the potential the coefficient is positive and if 
$|S_0| >
|\bar S_0|$ the logarithm is also positive.  We thus see that, under these
general conditions, the sign of the D term correction is determined solely by the
U(1)$_X$ charge 
$Q^X_a$.

Gauginos obtain no mass from this source.
 In addition, since the D term contribution to scalar masses (squarks, sleptons
and Higgs) is proportional to their U(1)$_X$ charge $Q^X_a$, in order to
suppress  flavor changing neutral current interactions the U(1)$_X$ charges of 
squarks and sleptons must be family independent.  
One interesting possibility is
that U(1)$_X$ is identified with the U(1) in E$_6$ which commutes
with SO(10).\footnote{The possibility of an anomalous U(1)$_X$ derived from
strings has recently been considered in the literature\cite{farragi}.  These
authors consider the same U(1)$_X$ as discussed here.} In this
case, the U(1)$_X$ charges on (16, 10, 1) in the 27 of E$_6$ are given by (1, -2, 4).
As a consequence, the standard model fermions, in sixteens of SO(10), have 
identical positive charge under this U(1)$_X$; thereby obtaining identical
positive mass squared corrections.\footnote{Note, $Tr Q^X > 0$, as assumed earlier, if 16s
dominate in the sum.}  Higgs, on the other hand, in tens of SO(10),
 have opposite charge and thus negative mass squared corrections. 

  The messengers generate SUSY breaking
masses for both scalars and gauginos.   Their contribution to scalar masses
are naturally family independent.

 \section{Discussion and Conclusions}

The model as it stands has one problem which is easily fixed.  The flavor
SU(2)$_F$
gauge symmetry is broken to U(1)$_F$ flavor via the vacuum expectation value of 
an SU(2)$_F$ vector $<S_3>$.  U(1)$_F$ remains unbroken at low energies and thus
there exists a massless gauge boson, a $\gamma_F$.   This $\gamma_F$ couples
to quarks and leptons through loops containing the messengers and standard
model gauge interactions.  A single $\gamma_F$ vertex with ordinary matter
is forbidden by U(1)$_F$ charge conjugation.  Nevertheless a two $\gamma_F$
coupling is allowed.   A new massless gauge interaction would probably have
been observed.  Thus we consider how to break this unwanted U(1)$_F$
symmetry.   This is in fact quite easy.  We can introduce SU(2)$_F$
doublets into the theory which naturally obtain vevs of order the
weak scale or larger.   Consider introducing $n_\phi$ doublet fields 
$\phi_\alpha, \; \bar \phi^\alpha$.   Assume for the moment they do not enter
the superspace potential $W$.  Thus at tree level they only enter the D term
for the SU(2)$_F$.\footnote{We assume they have zero U(1)$_X$
charge.}  They obtain mass at two loops due to SUSY breaking effects.  Recall
$S_3$ breaks both SU(2)$_F$ and SUSY.  As a result the massive gauge sector
in SU(2)$_F$/U(1)$_F$ acts as messengers of SUSY breaking for the $\phi$s.
Using the results of Giudice and Rattazzi\cite{giudice} (eqns. 61 - 65), we find
\begin{eqnarray}
m_\phi^2(S_0) & = 2 c' {\alpha_F^2 \over (4 \pi)^2} N [1 + r({N \over b'} - 1)]
|{F \over M_{mess}}|^2 &  
\end{eqnarray}
where $N = b' - b $; $b'$ and $b$ are the coefficients of the U(1)$_F$
and SU(2)$_F$ beta functions; $r = (c/c' - 1)/(b/b' - 1)$, and $c'$ and
$c$ are the quadratic Casimirs of the matter  gauge representations of the
groups U(1)$_F$ and SU(2)$_F$, respectively.  In our case $c' = 1/4$, $c = 3/4$,
$b' = - n_\phi$ and $b = 2 - n_\phi$ (see eqn. \ref{eq:b}).  
Thus $N = - 2$.   The mass squared correction to $\phi$ is negative
for  $n_\phi > 1$ and is given by
\begin{eqnarray}
m_\phi^2 & = - (n_\phi - 1) ({\alpha_F \over 4 \pi})^2 \;|{F \over M_{mess}}|^2
& . 
\end{eqnarray} 

Now consider  $n_\phi = 2$.  At the messenger scale $\phi, \; \bar \phi$ obtain
negative mass squared of order the weak scale.   The potential for $\phi , \; \bar \phi$ is
thus unbounded from below.   A contribution to the superspace potential of the form
\begin{eqnarray}
W & =  {1 \over M} \; (\phi_\alpha \; \bar \phi^\alpha)^2 &
\end{eqnarray}
leads to a potential for $\phi , \; \bar \phi$ of the form 
\begin{eqnarray}
V(\phi , \; \bar \phi) & = - |m_\phi^2| (|\phi|^2 + |\bar \phi|^2) + {2 \over M^2} \;
(|\phi|^2 + |\bar \phi|^2)\; |\phi_\alpha \; \bar \phi^\alpha|^2 &
\end{eqnarray}
and to a U(1)$_F$ breaking vev 
\begin{eqnarray}
<\phi>  & \sim  \sqrt{|m_\phi| \; M} & \sim 10^{9 - 10} \; \; \rm GeV
\end{eqnarray} 
for $M \sim 10^{16 - 18}$ GeV.
Note, with the additional $n_\phi = 2$ doublets, the SU(2)$_F$ beta function vanishes
at one loop.  The only effect this has on our previous results is that now $b = 0$ in (eqns.
\ref{eq:gammaS}, \ref{eq:dtermmass}) for the induced D term.

 We have presented a model for SUSY breaking which can provide a phenomenologically
interesting spectrum of sparticle masses.  Sparticles obtain both D term (eqns.
\ref{eq:gammaS}, \ref{eq:dtermmass}) and
gauge mediated messenger contributions (eqn. \ref{eq:mtilde}) to their mass.  
For phenomenological purposes our result suggests using the following form
for D term contributions
\begin{eqnarray}  \tilde m_a^2 &  = D \; M_2^2 \; Q^X_a &  \end{eqnarray}
where  $Q^X_a$ is the anomalous U(1)$_X$ charge of particle $a$, $M_2 =
{\alpha_2 \over 4 \pi} \; \Lambda$ (wino mass) can be taken to set the scale
for this contribution  and $D$ is an arbitrary parameter of order
one.  $Q^X_a$ is assumed to be family independent.  In the minimal model,
$Q^X_a = 1$ for quarks and leptons and $Q^X_a = -2$ for Higgs doublets.
The model has two fundamental scales.   The SUSY breaking scale  
$\sqrt{F_{S_3}} = \Lambda_s$ is the dynamical scale of a strong SU(2) gauge symmetry.  
The messenger scale $<S_3> = S_0$ is set by 
the Fayet-Iliopoulos D term.   In strings the latter scale is fixed
by the compactification scale.  The effective SUSY breaking scale in the
observable sector is given by $\Lambda = {2 F_{S_3} \over <S_3>} 
\approx 10^5$ GeV.  We assume that $S_0$ is bound from above by $\sim
10^{15}$ GeV in order to suppress gravity mediated SUSY breaking masses 
which are not gauranteed to be family independent.  The natural scale for $S_0$ (eqn.
\ref{eq:epsilon}) is of order $\sim M_{st}/10$.   Thus the scenario presented in
this paper would best fit into the strong coupling limit of the
heterotic string\cite{horava} in which $M_{st}$ is identified with the GUT scale
$\sim 10^{16}$ GeV.

A recent paper on the cosmological problems associated with gravitinos has found
extremely low reheat temperatures are necessary for weak scale gauge mediated
models with very light gravitinos \cite{cosmology}.   In our case, the gravitino mass is of order
12 GeV and the reheat temperature is much higher.  

Finally it has been argued that, with a particular mechanism for stabilizing the
dilaton in string theories, the dilaton SUSY breaking contribution always
dominates over the D term SUSY breaking contribution of an anomalous U(1)$_X$\cite{dilaton}.
If this is true, then the SUSY breaking contributions considered in this paper are
subdominant.  However, other mechanisms for stabilizing the dilaton may not have this
effect, see for example \cite{casas} where the dilaton is stabilized by
contributions to the superspace potential. In this case dilaton SUSY breaking vanishes.  
Clearly, this potential dilaton problem requires further study.

{\bf Acknowledgements}
Finally, this work is partially 
supported by DOE grant DOE/ER/01545-740.

%

\end{document}